\documentclass[aps,prl,twocolumn,showpacs,preprintnumbers,amsmath,amssymb,superscriptaddress]{revtex4}

\usepackage{graphicx}
\usepackage{psfrag}
\usepackage{color}

\newcommand{\un}[1]{\ensuremath{\mathrm{#1}}}

\begin{document}

\title{Dynamic Manipulation of Bose-Einstein Condensates With a
Spatial Light Modulator}

\author{V. Boyer}
\affiliation{Clarendon Laboratory, University of Oxford, Parks Road, Oxford,
OX1 3PU, UK}
\author{R. M. Godun}
\affiliation{Clarendon Laboratory, University of Oxford, Parks Road, Oxford,
OX1 3PU, UK}
\author{G. Smirne}
\affiliation{Clarendon Laboratory, University of Oxford, Parks Road, Oxford,
OX1 3PU, UK}
\author{D. Cassettari}
\affiliation{Clarendon Laboratory, University of Oxford, Parks Road, Oxford,
OX1 3PU, UK}
\author{C. M. Chandrashekar}
\affiliation{Clarendon Laboratory, University of Oxford, Parks Road, Oxford,
OX1 3PU, UK}
\author{A. B. Deb}
\affiliation{Clarendon Laboratory, University of Oxford, Parks Road, Oxford,
OX1 3PU, UK}
\author{Z. J. Laczik}
\affiliation{Department of Engineering Science, University of Oxford, Parks Road, Oxford,
OX1 3PJ, UK}
\author{C. J. Foot}
\affiliation{Clarendon Laboratory, University of Oxford, Parks Road, Oxford,
OX1 3PU, UK}

\date{\today}

\begin{abstract}

We manipulate a Bose-Einstein condensate using the optical trap
created by the diffraction of a laser beam on a fast ferro-electric
liquid crystal spatial light modulator.  The modulator acts as a phase
grating which can generate arbitrary diffraction patterns and be
rapidly reconfigured at rates up to 1~\un{kHz} to create smooth,
time-varying optical potentials.  The flexibility of the device is
demonstrated with our experimental results for splitting a
Bose-Einstein condensate and independently transporting the separate
parts of the atomic cloud.  \end{abstract}

\pacs{32.80.Pj, 42.40.My, 03.75.Kk}

\maketitle

The miniaturization of atomic traps is key to the use of ultracold
atoms in applications requiring full quantum control of the external degrees
of freedom, such as quantum information processing or matter wave
interferometry. As pointed out in Ref.~\cite{Lin2004}, quantum effects
on the propagation of the spatial wavefunction
arise when the trapping potential features details on a length scale
smaller than the atomic wavelength, typically $1~\mu m$ for ultracold
atoms.

One of the two main technologies to realize microtraps is the atom
chip~\cite{Reichel1999,Folman2002}, which consists of microfabricated
conducting wires able to generate highly configurable magnetic
traps.  
A common drawback is that the very proximity of
the wires or the surface holding the wires ($10 - 500~\un{\mu m}$)
results in undesired perturbations~\cite{Esteve2004,Jones2003,Lin2004}. 
Another limitation is that
magnetic trapping only works for a subset of all the Zeeman sublevels.

The alternative technology to magnetic atom chips is optical trapping.
Atoms in light fields far-detuned from atomic resonances can be
tightly confined in a non-dissipative trap, in any Zeeman sub-level of
the ground state~\cite{Miller1993} and far away from any physical
structure. Regular arrangements of microtraps, called
optical lattices~\cite{Porto2003}, are obtained by interfering 
intersecting laser
beams. More complex trapping structures have been
implemented on cold atoms using micro-lenses and other
micro-optic elements~\cite{Dumke2002a,Dumke2002}. However, compared 
to the magnetic atom chips, those techniques do not have much
flexibility built in, because the shape of the potential is contained
in a static optical setup, and only a small number of manipulations
can be performed~\cite{Dumke2002a}. 

Programmable diffractive
optics can efficiently extend the capabilities of micro-optic
components by generating arbitrary light patterns. In colloidal
physics
experiments using optical tweezers, it is common to generate multiple
traps by diffracting the light from spatial light modulators
(SLMs)~\cite{Grier2003}.
These programmable diffractive optical elements allow great
flexibility and dynamic control of the optical potentials, and dynamic
manipulation of plastic beads has been demonstrated~\cite{Sinclair2004}. 
However,
before the work reported here, the application of spatial light
modulation to move cold atoms had
been considered difficult because the motion of atoms is not
subject to damping (unlike samples in water), and variation of the
potential must be carried out extremely smoothly.  Previously, 
static intensity patterns have been proposed to create hollow guides
for cold atoms~\cite{McGloin2003}, and have been used to trap
atoms in multiple wells~\cite{Bergamini2004}.

We report
the manipulation of Bose-Einstein condensates with moving optical
tweezers generated by diffraction of a laser beam on the programmable 
phase grating
created by a ferro-electric liquid crystal SLM. We were able to split
the BEC into two or three pieces, and to move those pieces in a plane
perpendicular to the tweezers. This demonstrates the flexibility of this
technique and its ability to create a variety of dynamic optical traps 
for sub-microKelvin atoms.

An SLM is an array of pixels acting as individually tunable absorbers
or retardation waveplates which can imprint on a light beam a spatial 
amplitude
modulation or a spatial phase modulation respectively. Common SLMs are
made of nematic liquid crystal pixels whose birefringence is
controlled by polarizing them with an external electric field. They
exhibit a continuous phase retardation effect and a good diffraction
efficiency, and were successfully used to create static
patterns~\cite{Bergamini2004,Melville2003}.
However, their refresh rate, which is the rate at which the
retardation of a pixel can be changed from one value to another, is
of the order of 50 Hz, which is too low for most cold atom
applications. In order to generate a useful dynamic potential, the
device should be able to generate hundreds or thousands of different
patterns on the time scale of the experiment, typically a few seconds.
For this reason, we used a ferro-electric liquid crystal SLM, which
is much faster but is limited to binary values of the
retardation, resulting in an imprinted phase of 0 or
$\pi$~\cite{Hossack2003}.

\begin{figure}[tb]
\definecolor{gray}{rgb}{.6,.6,.6}
\begin{center}
\psfrag{Ei}{$E^\mathrm{in}$}
\psfrag{Eo1}{$E^\mathrm{out}_+$}
\psfrag{Eo2}{\color{black} $E^\mathrm{out}_-$}
\psfrag{Ed1}{$E^\mathrm{out}_0$}
\psfrag{Ed2}{\color{black} $E^\mathrm{out}_\pi$}
\psfrag{p}{$\pi$}
\psfrag{-22.5}{$-22.5^\circ$}
\psfrag{22.5}{$22.5^\circ$}
\psfrag{x}{$x$}
\psfrag{y}{$y$}
\includegraphics[width=.75\linewidth]{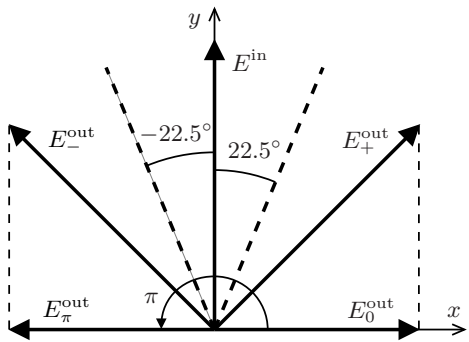}
\end{center}
\caption{\label{fig:rotation}
Polarization effects in a ferro-electric liquid crystal. The
electric field $E^\mathrm{in}$ of the incident light is polarized along the
$y$ axis. Depending on the two possible positions of the fast axis
(dashed lines), the output polarization $E^\mathrm{out}_\pm $is rotated by 
$\pm 45^\circ$.  The corresponding possible components along the $x$
axis $E^\mathrm{out}_0$ and $E^\mathrm{out}_\pi$ are
dephased with respect to each other by $\pi$. The component of the output polarization along the
$y$ axis is not dephased and goes into the zeroth order of the
diffraction pattern.}
\end{figure}

Our device is a 256$\times$256 square-pixel array of total size $4
\times 4$ mm from Displaytech, used in reflection. Let $x$ and $y$ be 
the main axes of
the chip. 
Ferro-electric liquid crystal materials are birefringent, with
two bi-stable orientations for the liquid crystal director. By
applying an appropriate electric field, the 
director can be switched between these two orientations, and
the fast axis can be rotated in the $xy$ plane between $\pm 22.5^\circ$
with respect to the $y$ axis (Fig.~\ref{fig:rotation}).
The retardation, which is a function of the
thickness of the material, is set to be $\lambda/2$ at $\lambda\simeq
700$~nm. As a result, the effect on light linearly polarized along $y$
is a rotation of the polarization through $\pm45^\circ$, depending on the sign
of the electric field, and the two possible values of the
$x$-component of the polarization of the rotated light are dephased by
$\pi$. Thus the $x$-component of the light reflected from the SLM has
a binary phase modulation and contributes to the diffraction pattern,
whereas the $y$-component of the light is not phase modulated and is
reflected into the zeroth order of the diffraction pattern.

We chose to use the diffraction pattern in the far field. Note that
the near field can also generate useful trapping potentials, 
as shown in Refs.~\cite{Newell2003,Schonbrun2005}.
In our case, in the scalar approximation, the diffracted electric 
field is the Fourier transform of the phase pattern on the SLM.
There is the same amount of light
in the plus and minus first orders, but in practice we only use the
plus first order.
This, together with the fact that the designed rotation of the incident
polarization is $45^\circ$ instead of $90^\circ$, that we use a 
wavelength for which
the retardation effect is less than $\lambda/2$, and that the pixels do
not join completely, leads to a diffraction efficiency
into the plus first order of only
3\%. However, this low efficiency is not of practical importance
because only $100~\un{\mu W}$ are 
required to trap a BEC in our experimental conditions.

The operation of a ferro-electric SLM presents two kinds of
problem.  First, in continuous operation, charge migration in the
liquid crystal reduces the retardation effect and the light power in
the diffraction pattern fades to zero with a time constant of 0.5~s.
Charge migration can be prevented by DC balancing the electric field
in such a way that the temporal average is close to zero. Thus the
problem is averted when the phase of each pixel is flipped regularly
in time. However, the second problem is that the switching of the state 
of a pixel takes about
250~$\mu s$, during which the birefringence is not well defined. 
Thus, flipping the phase of a large number of pixels at
the same time will produce a substantial flicker of the diffraction
pattern. Solving simultaneously the DC balancing and the flicker problems 
requires a compromise between opposite constraints on the level of
pixel changes between consecutive filters (phase modulation patterns) 
of a sequence.

\begin{figure}[tb]
\begin{center}
\psfrag{(a)}{(a) $\varphi=0$}
\psfrag{(b)}{(b) $\varphi=\pi/3$}
\psfrag{(c)}{(c) $\varphi=2\pi/3$}
\psfrag{(d)}{(d) $\varphi=\pi$}
\includegraphics[width=\linewidth]{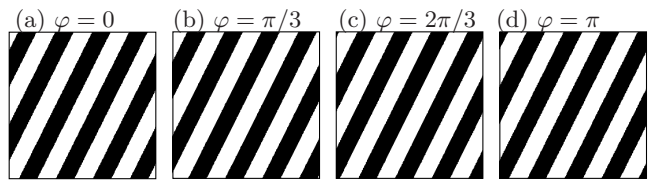}
\end{center}
\caption{\label{fig:glide}
Four different phase filters realizing the same diffraction
pattern consisting of a single spot. The pixels with a 0 phase are 
shown in white
and the pixels with a $\pi$ phase are in black. Each filter has the
phase pattern in a different position, corresponding to different
phases of the electric field, $\varphi$, at the spot
position.  The filters (a) and (d) are complementary.}
\end{figure}

Calculating the filter corresponding to a desired intensity
diffraction pattern (target) is a difficult problem in general. The
phase of the diffracted field, which is irrelevant to the atoms, is
unknown, and assuming a flat phase across the diffraction pattern and
taking the reverse Fourier transform of the electric field would give
a phase \emph{and} amplitude modulation diffraction grating. In order
to obtain a pure phase modulation grating, one has to use algorithms
such as iterative Fourier transform~\cite{Gerchberg1972}, 
genetic algorithm~\cite{Goldberg1989}, or
direct binary search (DBS)~\cite{Seldowitz1987}. We use a modified version of
DBS~\footnote{DBS is an iterative algorithm in which the target is
specified by providing the desired intensity level on a finite number
of points. The filter is randomly initialized, and the corresponding
diffracted intensity (current pattern) is calculated at the target
points. An error function provides a measurement of the distance
between the current pattern and the target. At each iteration, the
phase of a randomly chosen pixel of the filter is flipped if the
resulting variation in the error is negative} which takes care of the
flicker and DC balancing issues mentioned above.  As shown in
Ref.~\cite{Boyer2004a}, the flicker can be reduced by including a term in the 
DBS error function which keeps account of
the pixel changes between consecutive filters.  DC balancing requires
that \emph{all} the pixels are changed periodically in the sequence,
at least a few times per second. We enforce such a change by imposing
a time-varying value of the phase of the field at one of the target
points with an extra term in the error function. As shown in
Fig.~\ref{fig:glide}, modifying the overall phase of the field results
in changing the position of the phase pattern on the chip. A small
variation of the phase corresponds to the change of a small number of
pixels. By carefully setting the weight of the three terms in the
error function, we could generate sequences of filters which smoothly
moved a set of spots in the Fourier plane, while keeping the flicker
below a few percent of the total intensity, and preventing the
diffraction pattern from fading away.

To trap cold atoms in the optical potential created by the diffraction
pattern of the SLM, we initially create a BEC in a
magnetic trap. The BEC apparatus is similar to the one
described in Ref.~\cite{Arlt1999}, and consists of a pyramidal magneto-optical
trap (MOT) feeding a second MOT in a high vacuum glass cell. The atoms are
then transfered into a time-averaged orbiting
potential~\cite{Petrich1995} and are
evaporatively cooled until a BEC containing $2\times10^5$ atoms is
formed. In the final step, the magnetic trap is turned off while the
laser beam diffracted by the SLM (2~\un{mm} waist on the SLM) is 
turned on, transfering
the atoms into the optical potential. The diffraction pattern is imaged
on the trapping zone using the same optics as the absorption imaging
system. We thus have access to the atomic distribution in the
Fourier plane of the SLM. The smallest spot size waist we can create is $w =
3.8~\un{\mu m}$, slighly larger than the diffraction limit of the
optical system. At a wavelength $\lambda = 850$~nm and a power
$\mathcal{P} = 0.12$~mW, it creates an optical trap with oscillation
frequencies $1.2~\un{kHz}$ radially and $60~\un{Hz}$ in
the direction of propagation, and a
depth $\epsilon = 2~\un{\mu K}$~\footnote{The beam propagates
horizontally; gravity will reduce the trap depth
by about 25\%.}.  

A BEC of $2\times10^5$ atoms trapped in such a spot has dimensions of
$1.5~\un{\mu m} \times 1.5~\un{\mu m} \times 30~\un{\mu m}$ and a
chemical potential of $0.7~\un{\mu K}$, only one third of the trap
depth. This means that very few thermal atoms can be trapped, and the
BEC in equilibrium is almost pure. When heating is applied to the
cloud, fast evaporation and rethermalization bring the system back to
an equilibrium consisting of a quasi-pure BEC with fewer atoms.
Because we operate at high peak density, around
$10^{15}~\un{cm^{-3}}$, the collision rate is very high and this
rethermalisation happens on a timescale of a few milliseconds.
Unless the heating rate is very high, the cloud stays condensed and
the heating translates into atom losses, on top of the other sources of
loss. These are dominated by the 3-body recombination, which leads to
a measured loss rate per atom of $\sim 2~\un{s^{-1}}$ for a BEC of
$2\times10^5$ atoms.

\begin{figure}[htb]
\psfrag{0}{\color{white}{0~\un{ms}}}
\psfrag{100 ms}{\color{white}{100~\un{ms}}}
\psfrag{200 ms}{\color{white}{200~\un{ms}}}
\psfrag{?}{\color{white}{55~\un{\mu m}}}
\begin{center}
\includegraphics[width=0.4\linewidth]{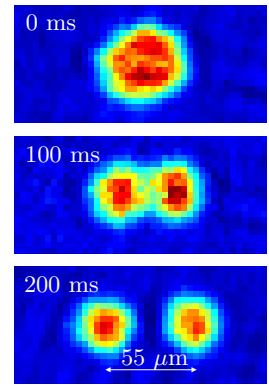}
\end{center}
\caption{\label{fig:split}(color online). 
Splitting and transport of a BEC with two optical tweezers. 
Each frame is
an absorption image of the atomic cloud, taken at the time shown on
the left.}
\end{figure}

The first of the two experiments that we report here splits the BEC into two
parts, as shown in Fig.~\ref{fig:split}. During $250~\un{ms}$, the BEC is
transfered into a light pattern consisting of two spots in the Fourier
plane separated by
$6.4~\un{\mu m}$~\footnote{Although this distance of $6.4~\un{\mu m}$ is
close to the spot size, the two spots are fully separated. Unlike
schemes relying on the incoherent addition of the potentials created
by two laser beams, for instance using an acousto-optic modulator
with multiple driving
frequencies~\cite{Shin2004a}, the ability to control the light phase at 
the target
allows increased versatility when generating a double well potential.
For example, it becomes possible to create two very closely spaced
spots which destructively interfere at the mid-point, allowing the
spots to be separated by a narrower potential barrier.}. 
After the
magnetic field is switched off, the two spots are moved away from each
other until they reach a separation of $55~\un{\mu m}$ after
$200~\un{ms}$. The refresh rate of the SLM is $500~\un{Hz}$ and the
displacement of the spots between each filter is $1/18$ of the
spot waist.
Figure~\ref{fig:split} shows the atomic spatial distribution measured
by absorption imaging, at various times during the splitting. Each
picture represents a different realization of the sequence for which
the separation process is interrupted at a given point by turning off the
trapping light. The atoms are allowed to free fall for $3~\un{ms}$
before the picture is taken.  During the imaging process, the atoms
are heated up by 
exchanging photons with a near resonant probe beam, resulting in atomic
clouds much more expanded than the original size in the trap. However
the distance between the clouds accurately reflects the separation
between the two traps. 

In an independent series of experiments, we accurately measure the
atom number by allowing the cloud to expand for $15~\un{ms}$ before taking
images, and we compare the atom loss of the moving sequence with that
of a static double well, which reflects the contribution of 3-body
recombination only. At the end of the separation sequence shown in
Fig.~\ref{fig:split},
the total atom loss is 50\%, half of which is due to the non-adiabatic
motion of the potential. The important result is that the non-adiabatic 
loss reduces to a few
percent when using smaller displacements of the spots between consecutive
frames of only $1/36$ of the spot waist.  We find that the splitting
generally works well for values of the refresh rate of the SLM ranging
from 200~\un{Hz} to 1~\un{kHz}, but some discrete values have to be avoided in
order to prevent strong heating of the BEC; such resonances occur
because the refresh rate can have harmonics resonant with the radial
oscillation frequency, resulting in a non-adiabatic linear motion
which can excite radial oscillations.

\begin{figure}[tb]
  \makebox[0pt]{\makebox[6ex]{}\raisebox{25ex}{\hspace*{2ex} \large \textbf{(a)}}}%
  \includegraphics[width=.47\linewidth, totalheight=9cm,
  keepaspectratio=true]{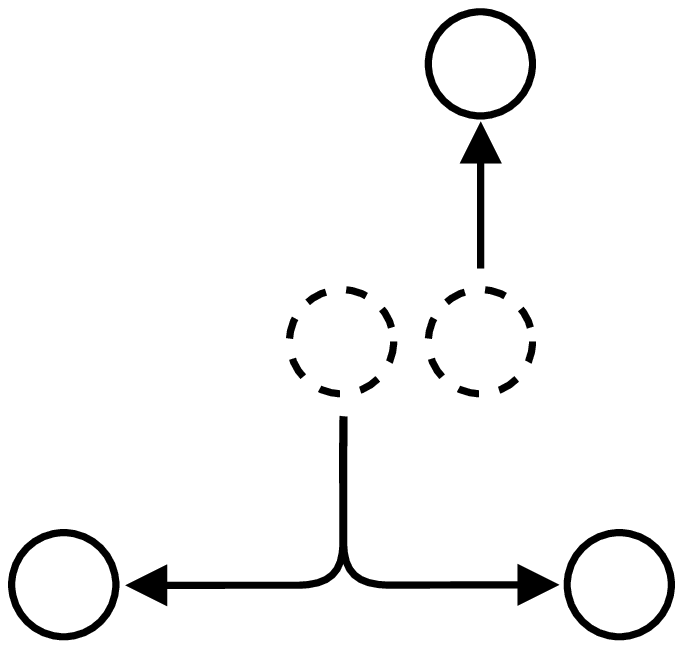}%
  \hfill%
  \psfrag{25 mum}{\color{white} {$25~\un{\mu m}$}}%
  \psfrag{(b)}{\color{white} \large \textbf{(b)}}%
  \includegraphics[width=.5\linewidth, totalheight=9cm,
  keepaspectratio=true]{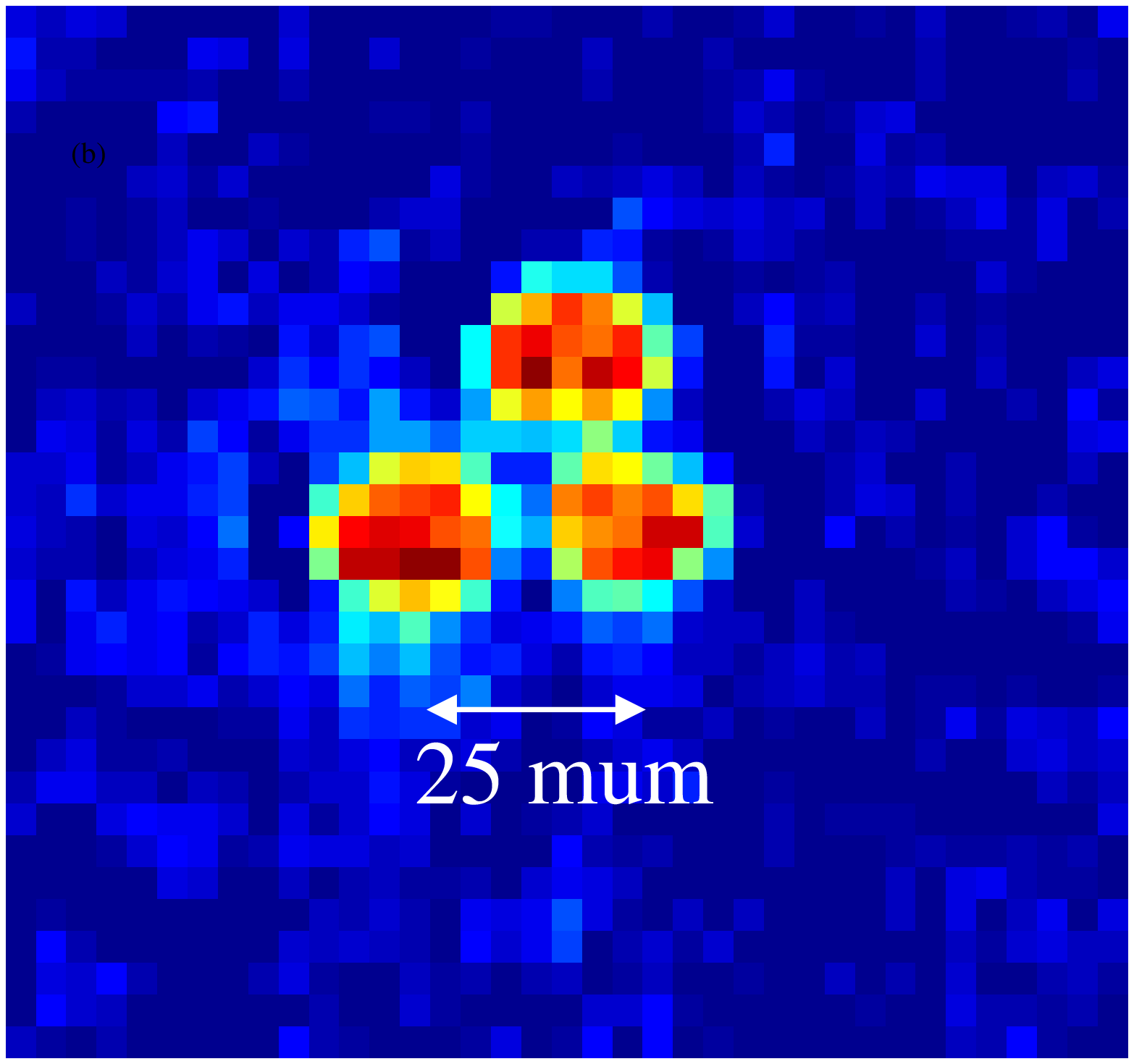}%
\caption{\label{fig:blobs}(color online).
Splitting of a BEC into 3 pieces. (a): Schematic of the dynamics of the potential, starting from a
double-well (dashed lines) to 3 separate traps (solid lines). (b):
Absorption imaging of the BEC in the final trap.}
\end{figure}

To demonstrate further the flexibility of our setup, we load the BEC
into a set of three spots in an arrangement that would not be easily
achievable using many other techniques. As shown in
Fig.~\ref{fig:blobs}(a), we start by loading the BEC directly into a
double-well and, initially, we move the two wells apart
vertically until they are separated by $20~\un{\mu m}$. We then split 
the lower spot into two spots which separate
horizontally to up to $25~\un{\mu m}$. The splitting required 
careful control of the
height and the shape of the potential barrier separating the newly
created pair of wells. A picture of the BEC in the final
3-spot arrangement is shown in Fig.~\ref{fig:blobs}(b). The length of
the sequence after the initial loading is 540~\un{ms}, and the refresh rate
of the SLM is set to 200~\un{Hz}. Because we are currently limited to only
128 different filters for technical reasons, we cannot
generate a motion of the potential as smooth as the one in the 2-spot
experiment. 
As a result,
a more severe heating leads to a final number of atoms which is half
the number expected when taking into account the 3-body
recombination only. However, using a larger number of filters would reduce
that heating.

In conclusion, we have shown that dynamic optical potentials created
with SLMs can be used to manipulate the external degrees of freedom of
cold atoms in a controlled manner and with little heating. We believe
that combined with high numerical aperture optics, resulting in
tightly confining optical tweezers, this versatile technique can play
a very important role for applications such as quantum information 
processing with neutral atoms, where one of the crucial steps is the
building of a bus able to carry individual qubits between memory and
processing units. Proposals using on-chip magnetic trapping to create quantum
processors with mobile qubits can be readily
adapted to diffractive optical trapping without having the drawbacks of
magnetic trapping. It should be emphasized that the optical traps
created by diffraction are not limited to two dimensions and can
extend in all directions.

So far, we
have focused on dynamic and non trivial arrangements
of spots, but generating other patterns, such as so-called doughnut modes, 
should be possible, for instance to study persistent
currents and superfluid properties.


\begin{acknowledgments}
We acknowledge support from EPSRC, EC (Marie-Curie fellowship, Cold Quantum
Gases network), the Royal Society,
and DARPA.
\end{acknowledgments}

\bibliography{paper}

\end{document}